# Percepta: High Performance Stream Processing at the Edge

Clarisse Sousa, Tiago Fonseca, Luis Lino Ferreira, Ricardo Venâncio, Ricardo Severino
*INESC-TEC/Instituto Superior de Engenharia do Porto*
Porto, Portugal
{cassa, calof, llf, ravrf, sev}@isep.ipp.pt

*Abstract*— The rise of real-time data and the proliferation of Internet of Things (IoT) devices have highlighted the limitations of cloud-centric solutions, particularly regarding latency, bandwidth, and privacy. These challenges have driven the growth of Edge Computing. Associated with IoT appears a set of other problems, like: data rate harmonization between multiple sources, protocol conversion, handling the loss of data and the integration with Artificial Intelligence (AI) models. This paper presents Percepta, a lightweight Data Stream Processing (DSP) system tailored to support AI workloads at the edge, with a particular focus on such as Reinforcement Learning (RL). It introduces specialized features such as reward function computation, data storage for model retraining, and real-time data preparation to support continuous decision-making. Additional functionalities include data normalization, harmonization across heterogeneous protocols and sampling rates, and robust handling of missing or incomplete data, making it well-suited for the challenges of edge-based AI deployment.

*Keywords—Data Stream Processing, Edge Computing, Internet of Things*

## I. Introduction

Cloud computing has become a dominant paradigm, centralizing data processing in high performance scalable infrastructures. However, as data volumes grew, traditional processing proved insufficient. Data Stream Processing (DSP) systems like Apache Spark, Heron, and Storm emerged to enable real-time data handling, transforming cloud-based processing [2-3].

The rise of the Internet of Things (IoT), with numerous distributed data-generating sources, exposed limitations of centralized clouds, particularly in latency, bandwidth, and privacy, thus leading to the advent of Edge Computing.

Unlike cloud infrastructures, Edge Computing processes data near its source, reducing latency, enhancing privacy, and supporting time-sensitive, resource-constrained applications like autonomous vehicles and healthcare [6]. Such applications demand rapid data processing, often within milliseconds, making centralized solutions impractical due to latencies. Edge Computing, while mitigating these issues and reducing costs, still requires efficient processing. Yet, most DSP systems, like Storm and Heron demand high computational resources not typically available at the edge [6-7].

Simultaneously, the integration of Artificial Intelligence (AI) with Edge Computing and IoT has introduced Edge AI [11], executing AI models directly on edge devices. Despite training challenges due to data volume requirements, they also rely on complete and correctly aggregated data during inference. For this reason, they can significantly benefit from DSP systems.

**Percepta**, the solution proposed in this paper, was initially created as part of the European project OPEVA (Optimization of Electric Vehicle Autonomy) [16], and can be classified as a DSP system with enhanced and specific capabilities for supporting AI models running at the edge, with a particular emphasis on Reinforcement Learning (RL) algorithms. In RL, an agent interacts with its environment by taking actions and receiving feedback in the form of rewards. Based on these rewards, the agent learns which actions lead to better outcomes, gradually improving its decision-making strategy over time. Percepta is designed to facilitate this process at the edge by computing reward functions directly from real-world interactions at each edge device and storing the necessary data for model retraining in the future, anonymizing it and delivering it to the node responsible for training the algorithms. Beyond these RL-specific features, Percepta also incorporates a broader set of capabilities essential for effective AI deployment at the edge, which also puts it apart from other DSP tools, including:

AI data conversion and preparation: Data collected from the environment is typically not in the format required for inference. As part of the system's processing, Percepta aggregates data and establishes relationships between different data sources, followed by normalization to ensure it can be effectively used by models. Once the inference is taken, decisions and set points are decoded and sent to the needed places.

Data rate harmonization and protocol conversion: Environmental data can originate from various sources using different communication protocols such as AMQP, MQTT or HTTP/S, with different reporting intervals and using different coding formats. For example, one device may send data every 5 minutes while another sends it once per hour. Percepta transparently handles these differences by aligning timestamps, aggregating or interpolating data when necessary. This harmonization ensures that the environmental data is delivered at the time resolution required by the RL algorithm.

Gap filling: Percepta also has to handle data loss, which is common in IoT applications where real-world data might be unavailable, sometimes simply because a specific sensor was turned off. Percepta is capable of detecting missing data and, when necessary, filling in the gaps to maintain the continuity and reliability of the input data and ensure the operation of the AI algorithms.

This paper is structured into the following sections: Section II introduces the concept of DSP and provides a comprehensive analysis of the state of the art in edge DSP. Section III details the architecture of Percepta, while Section IV explores a real-world use case where Percepta is currently being implemented. Additionally, Section V presents the conclusions.

## II. BACKGROUND AND RELATED WORK

As previously mentioned, a DSP system is designed to handle continuously generated data streams over time [8]. These systems apply user-defined logic to incoming raw data to extract value and make it immediately useful. Typically, these architectures adopt a graph-based model in which vertices represent sources (entities that produce data), operators (which apply processing logic to the data stream, such as aggregations, filtering, or transformations), and sinks (entities that consume the processed data), while edges represent the data streams that flow between the various operators [9].

Given this foundational model, there are several widely established DSP solutions in the market, and although they were designed to operate in distributed environments and clusters, it is important to address them as they often serve as a foundation for the development of DSP systems.

Apache Storm [10] is a general-purpose DSP system based on Spouts (data sources), Bolts (processing logic), Tuples (data units), and Topologies (data flow structure). It follows a master-slave architecture composed of two types of nodes: a Master Node, which runs the Nimbus daemon, responsible for managing topologies and assigning tasks to other nodes, and multiple Worker Nodes, which run a Supervisor daemon and Worker Processes. These Worker Processes receive tasks and spawn Executors, which are threads that apply the processing logic to the data. Apache Spark [10] is a DSP engine designed for fast and scalable big data analytics. It supports both batch and stream processing, offers high-level APIs, and performs in-memory computations, which significantly boost performance for iterative algorithms and real-time workloads. It is based on a modular architecture including components like Spark SQL, MLlib, GraphX, and Spark Streaming for real-time data streams. Apache Flink [10] is another open-source, stream-first distributed data processing framework known for its ability to process unbounded and bounded data streams with low latency and high throughput. Flink treats batch processing as a special case of streaming, offering powerful abstractions for event time processing, state management, and fault tolerance. Supports complex event processing, windowing, and exactly-once semantics, allowing its use for real-time analytics and reactive applications.

These are 3 examples of the most relevant DSP solutions, but several data stream processing systems have emerged to address varying performance and deployment requirements across cloud and edge environments. Heron, is a high-throughput, low-latency system optimized for cloud-scale deployments, offering improved scalability and resource isolation. In contrast, Apache Edgent targets constrained edge devices, providing a lightweight Java-based runtime designed to process data close to the source with minimal footprint, but it has seen limited maintenance and adoption in recent years. AgileDART [12] introduces a modular architecture aimed specifically at adaptive edge analytics, focusing on context-awareness and dynamic reconfiguration based on environmental changes, which is essential for mobility and fluctuating resources. EdgeWise [13] emphasizes secure and policy-driven stream processing, enabling fine-grained control over data flows across trusted and untrusted edge environments, with strong attention to privacy and compliance. Lastly, STEAM [14] is tailored for time-series data and industrial IoT contexts, optimizing for scalable deployment across heterogeneous edge nodes with a focus on interoperability and low overhead. These last four frameworks are indeed designed with edge environments in mind. However, none of them provide native support for RL inference, reward calculation and encoding and decoding logic and for data rate harmonization, protocol conversion and gap filling.

## III. PERCEPTA ARCHITECTURE

Percepta was created to efficiently fill these shortcomings by providing a lightweight and efficient stream processing solution tailored for edge computing scenarios, with built-in support for real-time RL inference.

This native support is especially important because, during inference, RL models require timely, reliable input data that must be accurately processed, filtered, and delivered in the correct format. Moreover, RL models depend on the continuous computation of rewards, which provide essential feedback to evaluate outcomes and guide future decisions. Without efficient data handling during inference, the model's responsiveness and overall performance can be severely impacted.

Percepta addresses these requirements by processing raw data in real-time, detecting anomalies such as data spikes, and replacing missing values based on historical patterns or recent observations. It also establishes meaningful relationships between different data streams according to the logic of each specific use case, ensures that the model receives updated and complete input data at regular intervals, and performs reward computations in real time, enabling immediate validation of model actions and supporting continuous learning and adaptation. All input data and model decisions are also logged in a database, enabling future analysis and potential retraining.

This section presents a detailed overview of Percepta's architecture (Figure 1), which is composed entirely of software components. It focuses on its core modules, support for multiple data providers and environments, and the flexible deployment model across both centralized and edge scenarios.

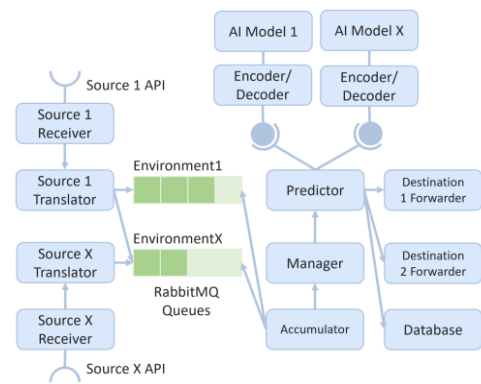

*Figure 1 - Architecture*

### A. Percepta Modules

**Receivers:** These are the system's input components. For each data source, there is a dedicated Receiver that adapts to the specific way the asset information is provided, whether it comes via MQTT, HTTP, AMQP, or other communication protocols.

**Translators:** Each data source also has an associated Translator that adjusts to the format of the incoming data, extracting only the relevant information. Whenever the Receiver obtains or requests data from the devices' APIs, it forwards the data to the Translator. The Translator then filters and structures the data into a standardized format and submits it to an internal RabbitMQ queue associated with the appropriate environment. In this context, an environment is an independent processing context with its own data flow, components, and AI model. For simplicity, the diagram shows a generic module instead of all environment-specific components (e.g., Managers, Accumulators).

**Accumulator:** The Accumulator component is agnostic to the data providers, i.e. it does not consider the origin of the data. Each environment has its own dedicated Accumulator instance, which listens to the corresponding RabbitMQ queue. Upon receiving a message, it forwards the data to the environment-specific Manager.

**Manager:** At the end of each time window (e.g., every 15 minutes), the Manager processes all the data collected during that period. It can prioritize the most recent entries, but it can also apply aggregation logic, such as calculating sums, averages, or other statistical operations, depending on the nature and requirements of each data source and type. If data is missing, the Manager predicts the missing values supported on historical data to ensure completeness of data, required by the AI algorithm. Once the processing period is finalized, the Manager analyzes the data to identify meaningful relationships within it. For instance, it may combine temperature readings from sensors of various brands within the same area to compute a weighted average that better reflects overall environmental conditions. Finally, the processed data is formatted and forwarded to the Predictor component.

**Predictor:** The Predictor component primary role is to route incoming data to the appropriate decision model associated with the environment, collect the resulting predictions, validate them, and compute the corresponding rewards. It then stores the input data, the decisions and computed rewards in a database for future analysis or model retraining and forwards the model decisions to the Forwarder components for processing or external delivery.

**Encoder/Decoder:** Different environments may be served by distinct AI models, each potentially requiring a unique data format. For each deployed model, an Encoder/Decoder component is implemented to translate the standardized format produced by the Manager into the specific format required by the model, which also depends on the characteristics of the environment. After inference, this component decodes the model's decisions back into a common format used across the system.

**Forwarders:** For each model decision destination, there is an associated Forwarder responsible for managing how the decisions are transmitted. As an example, consider a system responsible for controlling lighting, air conditioning, and security devices. If there is a smart light device that receives a "turn on" decision, then the decision is routed to the specific Forwarder associated with that system. This Forwarder ensures the decision is formatted and transmitted correctly.

*B. Handling Multiple Providers and Environments*

Percepta was designed with modular architecture, allowing for the integration of new data sources. To support an additional source, it is only necessary to implement a dedicated Receiver and Translator that handle the specific communication and data format of the new provider.

In addition to source-level modularity, the system also supports scalable multi-environment processing. This capability not only enables Percepta to be deployed at the edge serving a single environment but also allows cloud-based deployments that serve multiple isolated environments simultaneously. These environments operate independently, do not interfere with each other, and can utilize different AI models tailored to their needs. To achieve this isolation, each Receiver allocates a separate thread for every environment that requires data from that source. This design allows data collection to occur independently of the environment, even when sharing the same data providers.

Once data is received, it is passed to the Translator along with the environment identifier. The Translator extracts and formats the relevant information, then forwards it to the dedicated RabbitMQ queue for that environment. Each environment runs its own Accumulator thread listening to its queue, and upon receiving data, the Accumulator forwards it immediately to the corresponding Manager, which then forwards it to the Predictor.

*C. Deployment*

As mentioned in the previous section, although Percepta was specifically designed to operate at the edge, its architecture supports a wide range of deployment scenarios. It can run on edge devices serving a single local environment that may operate fully autonomously if relying only on local resources, even without internet connectivity, on fog environments serving multiple nearby environments, or entirely in the cloud, processing multiple environments simultaneously. This flexibility is enabled not only by Percepta's highly modular and multi-environment design but also by its distributed nature, facilitated through containerization, supported by Docker and Kubernetes.

IV. USE CASE: REAL-WORLD APPLICATION OF PERCEPTA

To better understand Percepta's purpose and how it operates, it is useful to examine the real-world challenge that led to its development.

The OPEVA project aims to accelerate the adoption of electric vehicles (EVs), leading to the development of an AI model (specifically, an RL model) designed to optimize the energy consumption of buildings, operating either at the edge, running locally on a controller within each building, or through a centralized approach [15].

To operate correctly, the AI model requires real-time periodic data from multiple sources, such as energy production from source X, consumption data from source Y, and EV charging data from source Z, which may be direct, for example, a device using Modbus, or indirect through third-party intermediaries, such as applications provided by the device manufacturers. In addition to data from specific assets within each installation, it may also rely on external sources such as energy price APIs, to determine the most cost-efficient times to consume energy, and weather APIs, to forecast renewable energy availability (e.g., solar or wind). Together, these data points build the real-time energy context of the facility.

However, these sources use a wide variety of protocols and formats to transmit data. Likewise, the method for delivering control commands to devices (such as turning on an appliance at a specific time) can also vary significantly. *Percepta* was developed not only to manage this communication heterogeneity, but also to support any type of AI model that consumes this data. Since each building may contain different assets, the version of the AI model deployed can vary from one installation to another.

In summary, once all *Percepta* components are configured according to the data sources, destinations, and the specific model used for a given building, or multiple models in a centralized setup, *Percepta* takes care of the entire data flow. At each interval, it aggregates the data, corrects anomalies, imputes missing values, encodes the input data in the format expected by the target AI model, receives the resulting commands, validates them, calculates rewards, stores both the input data and commands in a database, and forwards the latter to the appropriate endpoints.

Although still in its early stages, *Percepta* is already operational for testing purposes, both at the edge and in centralized configuration, alongside the AI model described earlier. This early deployment is particularly significant given that moving AI-based solutions from simulation to real-world environments is widely recognized as a complex and time-consuming process, often hampered by technical and operational hurdles [1]. In this context, while *Percepta* still requires more formal validation, it has so far shown strong potential as a practical solution for integrating RL models into real-world energy management systems.

## V. Conclusion and Future Work

This article presented Percepta, a lightweight and modular DSP solution that natively integrates data provisioning for RL models, addressing its key requirements, such as reward calculation for model validation and improvement. In addition to these capabilities, it also supports essential features such as timely data delivery from heterogeneous sources, error correction, and missing data imputation.

Beyond showcasing Percepta's edge-ready architecture, future work will involve validating its efficiency and robustness through systematic benchmarking. This will include evaluating network I/O performance under varying load conditions, measuring CPU, memory, and disk utilization across different stress levels, and assessing overall system performance across multiple deployment strategies, both centralized and decentralized, and on devices with diverse hardware capabilities. Future efforts will also focus on making the tool compatible with the Arrowhead Framework.

## Acknowledgments

This paper is supported by the OPEVA project that has received funding within the Chips Joint Undertaking (Chips JU) from the European Union's Horizon Europe Programme and the National Authorities (France, Czechia, Italy, Portugal, Turkey, Switzerland), under grant agreement 101097267. The paper is also supported by Arrowhead PVN, proposal 101097257. Views and opinions expressed are however those of the authors only and do not reflect those of the European Union or Chips JU. Neither the European Union nor the granting authority can be held responsible for them. The work in this paper is also partially financed by National Funds through the Portuguese funding agency, FCT - Fundação para a Ciência e a Tecnologia, within project LA/P/0063/2020. DOI10.54499/LA/P/0063/2020,https://doi.org/10.54499/LA/P/0063/2020, and through the FCT individual research grant 2024.00855.BD.